\documentclass[twocolumn,superscriptaddress,amssymb]{revtex4}

\usepackage{graphicx}

\def\result{-1240.9(3)}
\def\resultnoerr{-1240.9}

\def\resia{110.77155290(2)}
\def\resib{-134.983309(2)}
\def\resic{-18.1904(1)}
\def\resid{34.479723(1)}
\def\resie{42.93007(1)}
\def\resif{-159.43416(7)}
\def\resig{58.3492(2)}
\def\resih{142.936(2)}
\def\resii{23.4260(2)}
\def\resij{99.4005(4)}
\def\resik{16.86312(6)}

\begin{document}

\title{Four-Loop Collinear Anomalous Dimension in ${\mathcal N} = 4$
Yang-Mills Theory}

\author{Freddy Cachazo}

\affiliation{Perimeter Institute for Theoretical Physics,
Waterloo, Ontario N2J 2W9, Canada}

\author{Marcus Spradlin}

\affiliation{Brown University, Providence, Rhode Island 02912, USA}

\author{Anastasia Volovich}

\affiliation{Brown University, Providence, Rhode Island 02912, USA}

\begin{abstract}

We report a calculation in ${\cal N} = 4$ Yang-Mills of the
four-loop term $g^{(4)}$ in the collinear anomalous dimension
$g(\lambda)$ which governs the universal subleading infrared
structure of gluon scattering amplitudes. Using the method of
obstructions to extract this quantity from the $1/\epsilon$
singularity in the four-gluon iterative relation at four loops, we
find $g^{(4)} = \resultnoerr$ with an estimated numerical uncertainty of
$0.02\%$.
We also analyze the implication of our result for the
strong coupling behavior of $g(\lambda)$, finding support for
the string theory prediction computed recently by Alday and
Maldacena using AdS/CFT.

\end{abstract}

\maketitle

\section{Introduction}

Gluon scattering amplitudes in QCD and supersymmetric gauge theories
are notoriously difficult to compute, yet possess remarkably simple
hidden structure.  We expect the greatest simplicity in the
maximally supersymmetric ${\cal N} = 4$ Yang--Mills theory,
where planar $L$-loop amplitudes are believed to satisfy iterative
relations~\cite{Anastasiou:2003kj,Magnea:1990zb,Catani:1998bh,Sterman:2002qn,
Cachazo:2006tj,Bern:2006vw,Bern:2005iz}
in $L$.  These relations allow the complete, planar all-loop
$n$-particle MHV amplitude ${\mathcal A}$ to be written in an
exponential form due to Bern, Dixon and Smirnov~\cite{Bern:2005iz}.
Their ansatz for the $n=4$ particle
amplitude is
\begin{equation}
\label{eq:ansatz} \frac{\cal A}{{\cal A}_{\rm tree}}= ({\cal A}_{\rm
div}(s,t))^2 \exp\left[\frac{f(\lambda)}{8}
\log^2(t/s) + c(\lambda)\right],
\end{equation}
where $\lambda = g_{\rm YM}^2 N$ is the 't Hooft coupling, $s$ and $t$
are the usual four-particle Mandelstam invariants, and the infrared
divergences are encoded in the prefactors ${\cal A}_{\rm div}$.  In
dimensional regularization to $D = 4 - 2 \epsilon$ the divergences
take the form
\begin{eqnarray}
{\cal A}_{\rm div}(s,t) &=& \exp\Bigg[\left(- \frac{1}{8 \epsilon^2}
f^{(-2)}(\lambda \mu^{2 \epsilon}/s^\epsilon)\right.\cr
&&\qquad\left.- \frac{1}{4 \epsilon}
g^{(-1)}(\lambda \mu^{2 \epsilon}/s^\epsilon)
\right)+ (s \leftrightarrow t)\cr
&&\qquad +
  {\cal O}(\epsilon)\Bigg]
\end{eqnarray}
in terms of an IR cutoff scale $\mu$ (required on dimensional
grounds) and two functions $f^{(-2)}(\lambda)$ and
$g^{(-1)}(\lambda)$ which respectively are related to the
$f(\lambda)$ appearing in~(\ref{eq:ansatz}) and to a second function
called $g(\lambda)$ according to
\begin{equation}
f(\lambda)=\left(\lambda \frac{d}{d \lambda}\right)^2 f^{(-2)}(\lambda),
\quad
g(\lambda)=\lambda \frac{d}{d \lambda} g^{(-1)}(\lambda).
\end{equation}
The three functions $f(\lambda)$, $g(\lambda)$ and $c(\lambda)$ are
universal; the same functions appear in the exponential ansatz for
any planar $n$-particle MHV amplitude~\cite{Bern:2005iz}.

The function $f(\lambda)$ is well-known from another
role: it is
the cusp anomalous dimension, which also governs the scaling of
twist-two operators in the limit of large spin~$S$~\cite{Korchemsky:1988si,Korchemsky:1992xv,Gubser:2002tv,Kruczenski:2002fb},
\begin{equation}
\Delta\left( {\rm Tr}[Z D^S Z] \right)-S=f(\lambda) \log S +{\cal O}(S^0).
\end{equation}
At weak coupling it has been computed through four
loops~\cite{Moch:2004pa,Vogt:2004mw,Kotikov:2004er,Bern:2006ew,Cachazo:2006az},
\begin{eqnarray}
\label{eq:limitone}
&&f(\lambda) = 8 \left(\frac{\lambda}{16 \pi^2}\right)
- \frac{8\pi^2}{3} \left(\frac{\lambda}{16 \pi^2}\right)^2
+ \frac{88 \pi^4}{45} \left(\frac{\lambda}{16 \pi^2}\right)^3
\cr
&&
- \left(\frac{584 \pi^6}{315} -
64(1+r)\zeta_3^2\right) \left(\frac{\lambda}{16 \pi^2}\right)^4
+ {\cal O}(\lambda^5)
\end{eqnarray}
with $r=-2.00002(3)$ (the quantity in parentheses denotes the
current best
numerical uncertainty in the last digit), while AdS/CFT
calculations~\cite{Gubser:2002tv,Frolov:2002av} indicate the strong
coupling behavior
\begin{equation}
\label{eq:limittwo}
f(\lambda) = 4 \sqrt{\frac{\lambda}{16 \pi^2}}
- \frac{3 \log 2}{\pi} + {\cal O}(1/\sqrt{\lambda}).
\end{equation}
The recently proposed dressing phase~\cite{Beisert:2006ib,Beisert:2007hz}
for the asymptotic S-matrix~\cite{Beisert:2004hm,Arutyunov:2004vx,
Staudacher:2004tk,Beisert:2005fw,Beisert:2005tm,Eden:2006rx} of the
spin chain description of planar ${\cal N} = 4$
Yang-Mills 
implies that at finite $\lambda$, $f(\lambda)$ satisfies a certain
integral equation~\cite{Beisert:2006ez} whose solution is compatible with the
limits~(\ref{eq:limitone}),~(\ref{eq:limittwo})~\cite{Benna:2006nd,Alday:2007qf,Casteill:2007ct}
and moreover predicts
that $r = - 2$ in~(\ref{eq:limitone}).

Much less is known about the second function $g(\lambda)$
which governs the subleading infrared divergence and may be called
the ``collinear'' anomalous dimension~\cite{Korchemsky:1994is}.
Perturbative calculations through three
loops~\cite{Bern:2005iz} have established that
\begin{eqnarray}
\label{eq:glimitone}
g(\lambda)&=&
- 4 \zeta_3 \left(\frac{\lambda}{16 \pi^2}\right)^2
+ 8 (4 \zeta_5 + \frac{5}{9} \pi^2 \zeta_3)
\left(\frac{\lambda}{16 \pi^2}\right)^3
\cr
&&\qquad + g^{(4)}  \left(\frac{\lambda}{16 \pi^2}\right)^4
+ {\cal O}(\lambda^5),
\end{eqnarray}
and the purpose of this note is to report a numerical calculation
of the four-loop coefficient
\begin{equation}
\label{eq:result}
g^{(4)} = \result.
\end{equation}
It is an important outstanding problem to relate $g(\lambda)$ to
more familiar observables which could perhaps be computed using
integrability techniques.
In particular it would be interesting to derive an integral
equation satisfied by $g(\lambda)$.

An important step forward was recently taken by Alday and
Maldacena~\cite{Alday:2007hr}, who gave a prescription for computing
gluon amplitudes at strong coupling using AdS/CFT and found perfect
agreement with the ansatz~(\ref{eq:ansatz}).  A consequence of their
calculation is the prediction
\begin{equation}
\label{eq:strongcoupling}
g(\lambda) = 2 (1 - \log 2) \sqrt{\frac{\lambda}{16 \pi^2}} + {\cal O}(1)
\end{equation}
for the leading strong-coupling behavior of $g(\lambda)$. It is
important to mention that while $f(\lambda)$ is independent of the
IR renormalization scale, the same is not true for $g(\lambda)$.
Rescaling $\mu$ to $k \mu$ sends
$g(\lambda)$ to $g(\lambda)+2\log(k) f(\lambda)$.

In section II we review our calculational
method,
which is the same as the one used in~\cite{Cachazo:2006az} to
calculate $f^{(4)}$ to very high accuracy.
In section III we explore the connection between the
weak- and strong-coupling behavior of $g(\lambda)$.
First we use the Alday-Maldacena
result~(\ref{eq:strongcoupling}), together with the
known data~(\ref{eq:glimitone}) through three loops,
to build an interpolating function and make a `prediction' for
$g^{(4)}$.  Remarkably our result~(\ref{eq:result})
is within $7\%$ of this predicted value.  Finally we
use all currently available data on $g(\lambda)$ to
construct a Pad\'e approximant
which represents our best guess for the function's behavior at
finite $\lambda$.

\section{Method and Results}

To compute $g^{(4)}$
we apply the method of obstructions developed
in~\cite{Cachazo:2006mq,Cachazo:2006az}.
We begin with the four-loop iterative structure encoded
in~(\ref{eq:ansatz}), which in dimensional regularization takes
the form~\cite{Bern:2005iz}
\begin{widetext}
\begin{equation}
\label{eq:fone}
M^{(4)}(\epsilon) - M^{(3)}(\epsilon) M^{(1)}(\epsilon)
+ M^{(2)}(\epsilon)(M^{(1)}(\epsilon))^2
 - \frac{1}{2} (M^{(2)}(\epsilon))^2
- \frac{1}{4} (M^{(1)}(\epsilon))^4
= (f_0^{(4)} + f_1^{(4)} \epsilon) M^{(1)}(4 \epsilon)
+ {\cal O}(\epsilon^0),
\end{equation}
\end{widetext}
where $M^{(L)}(\epsilon)$
is the ratio of the $L$-loop amplitude to the tree amplitude
and $f_0^{(4)}$, $f_1^{(4)}$ are two numbers.
According to~\cite{Bern:2005iz}, whose conventions we follow except
as noted below, these are related to the four-loop coefficients $f^{(4)}$ and
$g^{(4)}$ in $f(\lambda)$ and $g(\lambda)$ by
\begin{equation}
\label{eq:dictionary}
f^{(4)}_0 = \frac{1}{2^4}\frac{1}{4} f^{(4)}, \qquad
f^{(4)}_1 = \frac{1}{2^4}\,2\,g^{(4)}.
\end{equation}
The factor of $1/2^4$ arises here at four loops because the
expansion parameter used by~\cite{Bern:2005iz} differs by a factor
of $2$ from that
of~\cite{Alday:2007hr} which we use in~(\ref{eq:limitone})
and~(\ref{eq:glimitone}).

Each of the five terms on the left-hand side of~(\ref{eq:fone}) is
separately an extremely complicated function of the ratio $x = t/s$
with singular behavior starting at ${\cal O}(1/\epsilon^8)$. It is
therefore rather remarkable that the five terms conspire to add up
to such a simple object involving only two free
constants~\cite{Magnea:1990zb,Catani:1998bh,Sterman:2002qn}. The
leading singularity of $M^{(1)}(4 \epsilon)$ on the right-hand side
is ${\cal O}(1/\epsilon^2)$, so we can determine $f^{(4)}_1$ by
reading off the coefficient of the $1/\epsilon$ pole on both sides
of~(\ref{eq:fone}).

\begin{figure}
\includegraphics{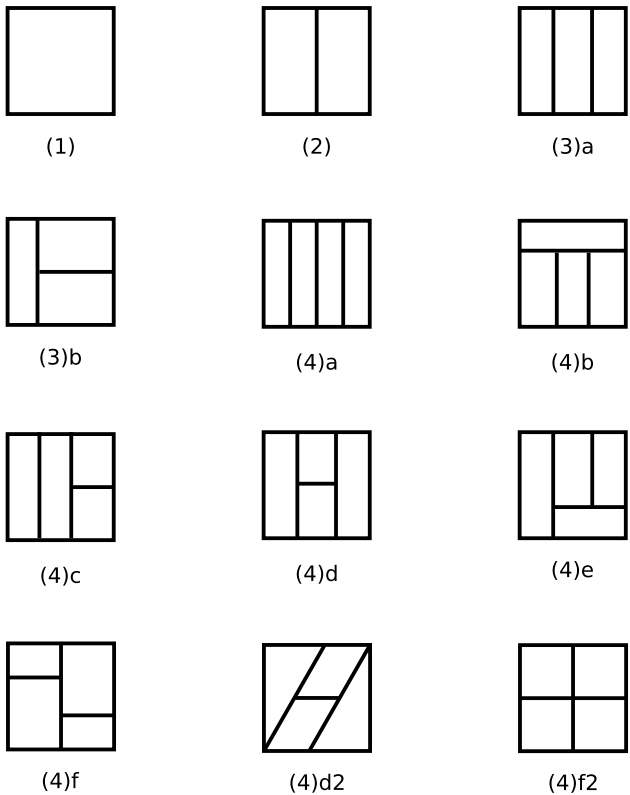}
\caption{The 12 integrals appearing in the $L \le 4$-loop four-particle
amplitudes in ${\mathcal N} = 4$ Yang-Mills theory.  We refer the
reader to~\cite{Bern:2006ew} for all necessary details.}
\end{figure}

The idea behind the method of obstructions,
explained in detail in~\cite{Cachazo:2006az}, is that we do not need to
fully compute all of the $M^{(L)}(\epsilon)$.  Rather, it is sufficient to
compute what are in some
sense the `constant pieces', since the $x$-dependent pieces are
guaranteed to cancel each other out on the left-hand side
of~(\ref{eq:fone}). This idea is made precise by writing each amplitude
as a Mellin transform, in the form
\begin{equation}
\int_{-i \infty}^{+i \infty} dy \ x^y F(y,\epsilon)
\end{equation}
for some $F(y,\epsilon)$.
It turns out that $f_1^{(4)}$ multiplies $\delta(y)$ in Mellin space,
so in order to isolate this term it is sufficient to read off
just the coefficient of $\delta(y)$ on both sides of~(\ref{eq:fone}),
throwing away the rest.
We call terms proportional to $\delta$-functions obstructions. The utility
of this method benefits significantly from the fact that we can work
directly in Mellin space, where it is relatively easy to construct
explicit formulas for Feynman
integrals~\cite{Smirnov:1999gc,Tausk:1999vh}.

It was further shown in~\cite{Cachazo:2006az} that obstructions obey
a product algebra structure, meaning that the obstruction in any
product of amplitudes is given by the product of their obstructions. This
statement is obvious in $y$-space where it hinges on the simple
fact that $\delta(y)$ convolved with itself is again $\delta(y)$.
Consequently we can calculate the obstruction $P^{(L)}(\epsilon)$
in each $M^{(L)}(\epsilon)$ separately and then insert those results
into the polynomial in~(\ref{eq:fone}).

The $L=1,2,3,4$ loop amplitudes we require may be expressed
in terms of the 12 scalar Feynman diagrams depicted
in Fig.~1~\cite{Green:1982sw,Bern:1997nh,Bern:2005iz,Bern:2006ew}.
The reader may find all necessary details
in~\cite{Bern:2005iz,Bern:2006ew}.
Our convention is that each
loop momentum integral comes with a factor
\begin{equation}
-\frac{1}{2} (-s)^{\epsilon/2} (-t)^{\epsilon/2}
\left[ -i \pi^{-D/2} e^{\epsilon \gamma}\right] \int d^D p.
\end{equation}
The standard convention includes only the factor in brackets.
The additional factor $(-s)^{\epsilon/2} (-t)^{\epsilon/2}$
renders all amplitudes
dimensionless (when the appropriate numerator
factors are included), but it does not
alter the form of~(\ref{eq:fone})
since each term in that equation has a common
factor $(-s)^{-2 \epsilon} (-t)^{-2 \epsilon}$.
The factor of $-1/2$ in front here eliminates the need for
a factor of $(-1/2)^L$ in~(\ref{eq:amplitudes}) below.

We have computed the obstructions in the 12 separate integrals
through the first eight orders in $\epsilon$ and present the results
in equation~(\ref{eq:obstructions}) at the end of the paper. The
calculation was performed using the algorithm explained
in~\cite{Cachazo:2006az} based on Czakon's {\tt MB}
program~\cite{Czakon:2005rk}.  Numerical integrations were performed
using {\sc CUBA}'s Cuhre algorithm~\cite{Hahn:2004fe}, and the digit
in parentheses denotes the reported uncertainty in the final digit.

An exact formula for ${\mathcal I}^{(1)}(\epsilon)$ was
given in~\cite{Cachazo:2006az}, as were the first
seven terms in the two- and three-loop amplitudes (though here
we indicate separately the contributions from the two three-loop
integrals).
No significant effort was spent on attempting to
improve the numerical accuracy of the ${\cal O}(1/\epsilon^3)$
and ${\cal O}(1/\epsilon^2)$ terms in the
individual four-loop integrals because these values play no role in this paper.

To find the full obstruction $P^{(L)}$ in the $L$-loop amplitude
we add together the contributions from the individual integrals
according to~\cite{Green:1982sw,Bern:1997nh,Bern:2005iz,Bern:2006ew}
\begin{eqnarray}
\label{eq:amplitudes}
P^{(1)} &=& {\mathcal I}^{(1)},\cr
P^{(2)} &=& 2 {\mathcal I}^{(2)},\cr
P^{(3)} &=& 2 {\mathcal I}^{(3)a} + 4 {\mathcal I}^{(3)b},\cr
P^{(4)} &=&
2 {\mathcal I}^{(4)a} + 4 {\mathcal I}^{(4) b}
+ 4 {\mathcal I}^{(4)c} + 2 {\mathcal I}^{(4)d} + 8 {\mathcal I}^{(4) e}
\cr
&&\qquad
+ 4 {\mathcal I}^{(4) f} - 4 {\mathcal I}^{(4) d2} - {\mathcal I}^{(4) f2}.
\end{eqnarray}
Plugging the resulting expressions for $P^{(L)}$ into~(\ref{eq:fone})
in place of $M^{(L)}$ and using the relation~(\ref{eq:dictionary})
leads to the advertised value~(\ref{eq:result}).

\section{Bridging Weak and Strong Coupling}

As mentioned above, string theory provides predictions for how
$f(\lambda)$ and $g(\lambda)$ behave at strong coupling. It is
therefore tempting to try to make a connection between the weak
and strong coupling regimes. An
approximation scheme that has been very successful for $f(\lambda)$
is the following. If $n+1$ orders of the expansion around
$\lambda=0$ are known, then one constructs an approximating function
$\hat f(\lambda)$ as an appropriate solution to the polynomial
equation~\cite{Kotikov:2003fb}
\begin{equation}
\label{eq:approx} \left(\frac{\lambda}{16\pi^2}\right)^n =
\sum_{r=n}^{2n} c_r \hat f(\lambda)^r,
\end{equation}
where the coefficients $c_r$ are determined by imposing that $\hat
f(\lambda)$ agree with $f(\lambda)$ through ${\cal O}(\lambda^{n+1})$.

The form of~(\ref{eq:approx}) incorporates a strong
coupling expansion for $\hat f(\lambda)$ in the parameter
$4 \pi/\sqrt{\lambda}$ with leading term $\alpha \sqrt{\lambda}/4 \pi$
where $\alpha^{2 n} = 1/c_{2n}$.

Using the expansion of $f(\lambda)$ up to four loops one finds an
approximating function $\hat f(\lambda)$ that agrees, within a few percent,
with the string theory strong coupling prediction and with the BES
ansatz~\cite{Beisert:2006ez} for all positive values of the coupling.

The success of this approximation scheme makes its application to
$g(\lambda)$ a natural thing to attempt. However, since $g(\lambda)$
is negative for small $\lambda$ but becomes positive at strong
coupling, it must have at least one zero for $\lambda > 0$.
Clearly, any approximating
function of the form~(\ref{eq:approx}) (with an appropriate
modification because $g(\lambda)$ only starts at two loops)
\begin{equation}
\label{eq:newapprox} \left(\frac{\lambda}{16\pi^2}\right)^{2n} =
\sum_{r=n}^{4n} c_r \hat g(\lambda)^r,
\end{equation}
leads to a $\hat g(\lambda)$ that cannot possibly have a zero for
$\lambda\neq 0$ and hence a contradiction.
One way to find a consistent approximating function was introduced
in~\cite{Alday:2007hr}. The idea is to perform a change of the IR
renormalization scale $\mu\to \exp(\xi/2)\mu$. Then, as discussed 
above, $g(\lambda)\to g(\lambda)+\xi f(\lambda)$.
The new function $\tilde g(\lambda ,\xi) =g(\lambda)+\xi f(\lambda)$
can now be approximated by using~(\ref{eq:approx}) for any $\xi>0$.

Consider the approximation~(\ref{eq:approx}) for
$\hat{\tilde g}(\lambda, \xi)$ with $n=2$,
\begin{equation}
\label{eq:approxG} \left(\frac{\lambda}{16\pi^2}\right)^2 = c_2 \hat
{\tilde g}(\lambda,\xi)^2+ c_3 \hat {\tilde g}(\lambda,\xi)^3+ c_4
\hat {\tilde g}(\lambda,\xi)^4.
\end{equation}
This equation depends on four parameters, namely, $c_2,c_3,c_4$ and
$\xi$. Imposing that the approximating function $\hat{\tilde
g}(\lambda,\xi)$ agrees with ${\tilde g}(\lambda,\xi)$ through
${\cal O}(\lambda^3)$ determines the three coefficients
$c_2,c_3,c_4$ in terms of $\xi$. In~\cite{Alday:2007hr} the
resulting approximating function was extrapolated to strong coupling
and compared to the string theory prediction for a range of $\xi$.
At the special value $\xi =  \frac{1}{2} \log 2$, for example, the
approximation gives $\hat{\tilde g}(\lambda) = 1.37 \sqrt{\lambda}/4
\pi$ compared to the string theory prediction $\tilde g(\lambda) = 2
\sqrt{\lambda}/4 \pi$.

Here we would like to use the results of Alday and Maldacena to
make a prediction for $g^{(4)}$ to compare our result~(\ref{eq:result}) to.
Imposing that $\hat{\tilde g}$ agrees with the string theory prediction
at strong coupling fixes the parameter
$\xi \approx 0.73679$.  Then
all four parameters $c_2, c_3, c_4$ and $\xi$ are completely
fixed and we can expand the resulting $\hat{\tilde g}$ in
$\lambda$ to ${\cal O}(\lambda^4)$
to find the predicted value $g^{(4)} \approx -1336.9$.
Remarkably our result~(\ref{eq:result}) is only $7\%$ away from this value.
We interpret this as good evidence for the string theory strong
coupling prediction of Alday and Maldacena~\cite{Alday:2007hr}.

Finally we present also a more conventional Pad\'e approximant
for $g(\lambda)$ based upon all currently available data.
Here we follow the approach described in~\cite{Bern:2006ew} where a
$[3/2]$ Pad\'e approximant in terms of the auxiliary variable $u =
\sqrt{1 + \lambda/\pi^2}$ was considered for $f(\lambda)$.
Specifically we consider
the ansatz
\begin{equation}
\label{eq:padeansatz} G(\lambda) = (u - 1)^2 \frac{N_0 + N_1 u}{1 +
D_1 u + D_2 u^2}
\end{equation}
with four parameters
$N_0,N_1,D_1,D_2$.
The choice $u = \sqrt{1 + \lambda/\pi^2}$ was
motivated by evidence that $f(\lambda)$ should have
a branch point at $\lambda = - \pi^2$.
It seems reasonable to guess that the same might be
true for $g(\lambda)$, though we have no direct evidence for this guess.

\begin{figure}
\includegraphics[scale=0.75]{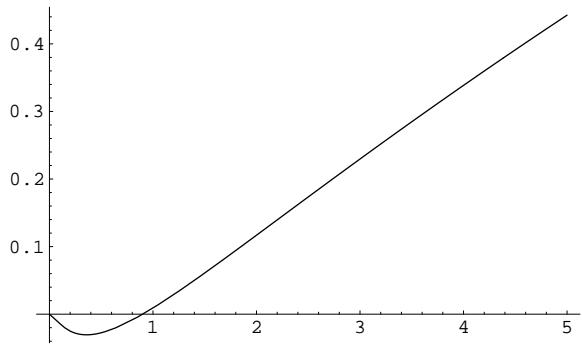}
\caption{A Pad\'e approximant for the
function $g(\lambda)$ versus $\lambda/16 \pi^2$, obtained by
fitting the ansatz~(\ref{eq:padeansatz}) to
all available data at weak and strong coupling.}
\end{figure}

The four parameters in~(\ref{eq:padeansatz}) can be uniquely
determined by fitting to all available data---the perturbative
expansion through four loops as well as the strong-coupling
limit~(\ref{eq:strongcoupling}).
The resulting approximant, displayed in Fig.~2, may be considered
our best candidate picture of $g(\lambda)$ at the moment.  Note that it
is minimal in the sense that $G(\lambda)$ was designed to have
a single zero along the positive real axis, whereas the true $g(\lambda)$
could potentially have any odd number of zeros.
Curiously, the zero lies very close
to $\lambda/16 \pi^2 = 1$, although it is impossible to conclude
based on the limited available data whether or not this is just
a coincidence.

\vfil

\begin{acknowledgments}

We have benefited from discussions with Z.~Bern, L.~Dixon and J.~Maldacena.
The research of FC at the Perimeter Institute is supported in part by
funds from NSERC of Canada and MEDT of Ontario. The research of
MS is supported by NSF grant PHY-0610259 and by an OJI award
under DOE grant DE-FG02-91ER40688.
The research of AV is supported by NSF CAREER Award PHY-0643150.
This work was made possible by the facilities of the Shared Hierarchical
Academic Research Computing Network (SHARCNET:{\tt www.sharcnet.ca}).

\end{acknowledgments}

\begin{widetext}
\begin{eqnarray}
\label{eq:obstructions}
{\mathcal I}^{(1)}(\epsilon) &=&
- \frac{2}{\epsilon^2}
+ \frac{2 \pi^2}{3}
+ \frac{17 \zeta_3}{3} \epsilon
+ \frac{41 \pi^4}{720} \epsilon^2
- \left[\frac{59 \pi^2 \zeta_3}{36} - \frac{67 \zeta_5}{5} \right] \epsilon^3
- \left[ \frac{\pi^6}{4320} + \frac{70 \zeta_3^2}{9} \right] \epsilon^4
\cr
&&\qquad
- \left[ \frac{143 \pi^4 \zeta_3}{864} + \frac{253 \pi^2 \zeta_5}{60} - \frac{261 \zeta_7}{7} \right] \epsilon^5
+ {\cal O}(\epsilon^6),
\cr
{\mathcal I}^{(2)}(\epsilon) &=&
+ \frac{1}{\epsilon^4}
- \frac{5 \pi^2}{8} \frac{1}{\epsilon^2}
- \frac{65 \zeta_3}{12} \frac{1}{\epsilon}
- \frac{\pi^4}{180}
+ \left[ \frac{77 \pi^2 \zeta_3}{24} - \frac{463 \zeta_5}{20} \right] \epsilon
- \left[ \frac{1999 \pi^6}{60480} - \frac{95 \zeta_3^2}{36} \right] \epsilon^2
\cr
&&\qquad
+ \left[ \resia \right] \epsilon^3
+ {\cal O}(\epsilon^4),
\cr
{\mathcal I}^{(3)a}(\epsilon) &=&
- \frac{2}{9} \frac{1}{\epsilon^6}
+ \frac{3 \pi^2}{16} \frac{1}{\epsilon^4}
+ \frac{131 \zeta_3}{72} \frac{1}{\epsilon^3}
+ \frac{187 \pi^4}{8640} \frac{1}{\epsilon^2}
- \left[ \frac{57 \pi^2 \zeta_3}{32} - \frac{841 \zeta_5}{120} \right] \frac{1}{\epsilon}
\cr
&&\qquad
+ \left[\frac{527479 \pi^6}{4354560} + \frac{265 \zeta_3^2}{72}\right]
+ \left[\resib \right] \epsilon
+ {\cal O}(\epsilon^2),
\cr
{\mathcal I}^{(3)b}(\epsilon) &=&
- \frac{2}{9} \frac{1}{\epsilon^6}
+ \frac{19 \pi^2}{96} \frac{1}{\epsilon^4}
+ \frac{241 \zeta_3}{144} \frac{1}{\epsilon^3}
- \frac{241 \pi^4}{10368} \frac{1}{\epsilon^2}
- \left[ \frac{2321 \pi^2 \zeta_3}{1728} - \frac{1009 \zeta_5}{80} \right] \frac{1}{\epsilon}
\cr
&&\qquad
- \left[\frac{605393 \pi^6}{26127360} + \frac{17 \zeta_3^2}{48}\right]
+ \left[\resic \right] \epsilon + {\cal O}(\epsilon^2),
\cr
{\mathcal I}^{(4)a}(\epsilon) &=&
+ \frac{1}{36} \frac{1}{\epsilon^8}
- \frac{187 \pi^2}{6912} \frac{1}{\epsilon^6}
- \frac{1169 \zeta_3}{3456} \frac{1}{\epsilon^5}
- \frac{277 \pi^4}{25920} \frac{1}{\epsilon^4}
+ 6.204418(2) \frac{1}{\epsilon^3}
- 63.9795(1) \frac{1}{\epsilon^2}
+ \left[ \resid \right] \frac{1}{\epsilon} + {\cal O}(1),
\cr
{\mathcal I}^{(4)b} (\epsilon) &=&
+ \frac{1}{36} \frac{1}{\epsilon^8}
- \frac{211 \pi^2}{6912} \frac{1}{\epsilon^6}
- \frac{601 \zeta_3}{1728} \frac{1}{\epsilon^5}
+ \frac{1181 \pi^4}{414720} \frac{1}{\epsilon^4}
+ 7.7902(1) \frac{1}{\epsilon^3}
- 14.757(2) \frac{1}{\epsilon^2}
+ \left[ \resif \right] \frac{1}{\epsilon} + {\cal O}(1),
\cr
{\mathcal I}^{(4)c}(\epsilon) &=&
+ \frac{1}{36} \frac{1}{\epsilon^8}
- \frac{29 \pi^2}{864} \frac{1}{\epsilon^6}
- \frac{1175 \zeta_3}{3456} \frac{1}{\epsilon^5}
+ \frac{3721 \pi^4}{414720} \frac{1}{\epsilon^4}
+ 8.31668(1) \frac{1}{\epsilon^3}
-  22.705(2) \frac{1}{\epsilon^2}
+ \left[ \resie \right] \frac{1}{\epsilon} + {\cal O}(1),
\cr
{\mathcal I}^{(4)d} (\epsilon) &=&
+ \frac{1}{36} \frac{1}{\epsilon^8}
- \frac{169 \pi^2}{6912} \frac{1}{\epsilon^6}
- \frac{521 \zeta_3}{3456} \frac{1}{\epsilon^5}
- \frac{7 \pi^4}{720} \frac{1}{\epsilon^4}
- 11.1550(2) \frac{1}{\epsilon^3}
+ 12.30(1) \frac{1}{\epsilon^2}
+ \left[ \resih \right] \frac{1}{\epsilon} + {\cal O}(1),
\cr
{\mathcal I}^{(4)e} (\epsilon) &=&
+ \frac{1}{36} \frac{1}{\epsilon^8}
- \frac{49 \pi^2}{1728} \frac{1}{\epsilon^6}
- \frac{1657 \zeta_3}{6912} \frac{1}{\epsilon^5}
- \frac{1441 \pi^4}{829440} \frac{1}{\epsilon^4}
- 2.4497(2) \frac{1}{\epsilon^3}
+ 7.081(6) \frac{1}{\epsilon^2}
+ \left[ \resig \right] \frac{1}{\epsilon} + {\cal O}(1),
\cr
{\mathcal I}^{(4)f} (\epsilon) &=&
+ \frac{1}{18} \frac{1}{\epsilon^8}
- \frac{235 \pi^2}{3456} \frac{1}{\epsilon^6}
- \frac{1001 \zeta_3}{1728} \frac{1}{\epsilon^5}
+ \frac{6497 \pi^4}{207360} \frac{1}{\epsilon^4}
+ 4.65725(7) \frac{1}{\epsilon^3}
+ 18.344(2) \frac{1}{\epsilon^2}
+ \left[ \resii \right] \frac{1}{\epsilon} + {\cal O}(1),
\cr
{\mathcal I}^{(4)d2} (\epsilon) &=&
- \frac{\zeta_3}{24} \frac{1}{\epsilon^5}
+ \frac{11 \pi^4}{6912} \frac{1}{\epsilon^4}
+ 2.2648(2) \frac{1}{\epsilon^3}
- 0.943(2) \frac{1}{\epsilon^2}
+ \left[ \resik \right] \frac{1}{\epsilon} + {\cal O}(1),
\cr
{\mathcal I}^{(4)f2} (\epsilon) &=&
+ \frac{1}{9} \frac{1}{\epsilon^8}
- \frac{235 \pi^2}{1728} \frac{1}{\epsilon^6}
- \frac{1073 \zeta_3}{864} \frac{1}{\epsilon^5}
+ \frac{6467 \pi^4}{103680} \frac{1}{\epsilon^4}
+ 14.5224(3) \frac{1}{\epsilon^3}
+ 54.854(2) \frac{1}{\epsilon^2}
+ \left[ \resij \right] \frac{1}{\epsilon} + {\cal O}(1).
\end{eqnarray}
\end{widetext}

\end{document}